\documentclass[preprint,showpacs,aps,preprintnumbers,amsmath,amssymb]{revtex4-1}

\usepackage{graphicx,epsfig}
\usepackage{dcolumn}
\usepackage{bm}
\newcommand{\be}{\begin{equation}}
\newcommand{\ee}{\end{equation}}
\newcommand{\bse}{\begin{subequations}}
\newcommand{\ese}{\end{subequations}}
\newcommand{\bary}{\begin{eqnarray}}
\newcommand{\eary}{\end{eqnarray}}
\newcommand{\bwt}{\begin{widetext}}
\newcommand{\ewt}{\end{widetext}}

\begin{document}


\title{On the non-detection of Glashow resonance in IceCube}
\author{Sarira Sahu$^{a,b}$ }
\email{sarira@nucleares.unam.mx}
\author{Bing Zhang$^{c}$}
\email{zhang@physics.unlv.edu}

\affiliation{$^{a}$Instituto de Ciencias Nucleares, Universidad Nacional Aut\'onoma de M\'exico, 
Circuito Exterior, C.U., A. Postal 70-543, 04510 Mexico DF, Mexico}
\affiliation{$^{b}$Astrophysical Big Bang Laboratory, RIKEN, Hirosawa, Wako, Saitama 351-0198, Japan}
\affiliation{$^{c}$Department of Physics and Astronomy, University of
  Nevada, Las Vegas, NV 89154, USA}


\begin{abstract}
Electron anti-neutrinos at the Glashow resonance (GR, at $E_{\bar \nu_e} \sim
6.3$ PeV) have an enhanced probability to be detected. With three
neutrinos detected by IceCube in the (1-2) PeV energy range at
present, one would expect that about 1 to 4  GR $\bar\nu_e$ should have been detected. The high-energy $\sim 8.7$ PeV muon neutrino detected by IceCube
may not be a GR event. If so, we expect to detect 50 to 70  GR $\bar\nu_e$,
then one would have a ``missing Glashow-resonance problem''. This would suggest (1) that 
$p\gamma$ interaction rather than $pp$ interaction is the dominant
channel to produce the observed IceCube high-energy neutrinos; (2) that
multi-pion $p\gamma$ interactions are suppressed; and (3) 
that the magnetic field and photon energy density in the $p\gamma$
emission region is such that significant $\mu^+$ cooling occurs before
decaying, yet $\pi^+$'s essentially do not cool before decaying.

\end{abstract}

\maketitle

\section{Introduction}
\label{sec:intro}

Observations of TeV-PeV neutrino events by
the IceCube Neutrino Observatory \cite{Aartsen:2013jdh,Aartsen:2013bka} 
opened up the window to study
the astrophysical origin of high energy neutrinos and their production
mechanisms. IceCube detects shower and track events from the deep
inelastic scattering of high energy neutrinos off the nucleons  in the
ice. These high energy neutrinos are produced in the astrophysical
environment due to the interaction of high energy cosmic rays with the
background gas ($pp$) or radiation ($p\gamma$) to produce pions. The
subsequent decay of charged pions and muons would produce high energy
neutrinos detected by IceCube. 

The interaction cross sections of high energy neutrinos and
anti-neutrinos with atomic electrons are very small compared to those
of the interactions with nucleons. However, in the resonant scattering
of 
\be
{\bar\nu}_ee^-\rightarrow W^-\rightarrow {\rm ~anything ~(hadrons + leptons)},
\ee
 the electron anti-neutrinos of energy
 \be
E_{\bar \nu}\simeq M^2_W/(2 m_e)\simeq 6.3 ~{\rm PeV}
\ee
have an enhanced probability to interact with the atomic electrons in
the ice to produce the on-shell $W^-$ boson. This is the so called
{\it Glashow resonance} (GR)\cite{Glashow:1960zz}. The cross section
at GR is about 300 times higher than that of the charged
current (CC) neutrino-nucleon interaction. 

This process is unique and of particular interest 
because of the dramatic increase in the event rate at the resonance
energy. At this energy the $W^-$ boson will predominantly decay to
hadrons (68\%). Other channels include $W^-$ boson decay to three
species of charged leptons and their
corresponding anti-neutrinos, respectively. Each of the leptonic channel has a branching ratio of $\sim 11\%$. So
the fraction of the leptonic decay mode which produces a muon and its
associated track-like event is small. On the other hand, the cascade/shower
from the hadronic decay is the most promising way of detecting the
GR. As the Earth is opaque to very high energy neutrinos, only the 
downward to horizontal going GR events can be observed by IceCube\cite{Barger:2012mz,Barger:2014iua}.

The showers due to the decay of resonantly produced W-bosons can have
peaks at three different energies depending on its decay
channel \cite{Kistler:2013my,Anchordoqui:2014hua,Biehl:2016psj}. The dominant one is at
6.3 PeV which 
is produced due to hadronic decay of the W-boson. If the W-boson
decays to $e^-$ and ${\bar\nu_e}$ ($W^-\rightarrow e^-{\bar\nu_e}$), a
peak at 3.2 PeV is formed. A third peak can be
formed at 1.6 PeV from the $\tau$ decay mode ($\tau^-\rightarrow
e^-{\bar\nu_e}{\nu_{\tau}}$). The decay to muon will give a track
without any peak.

The energy estimate from a cascade event in IceCube is more accurate
compared to a muon track. Due to the much higher interaction cross
section of the GR process, it is expected to contribute substantially to the event rate in IceCube.
This would offer a good chance to detect a signal from electron anti-neutrinos
of astrophysical origin, and would provide information about the
production mechanisms of high energy neutrinos in the astrophysical sources.

So far IceCube has detected more than 80 events that might have an
astrophysical origin, out of which three events have the
highest energy in the range of (1-2) PeV and all of them are
shower events\cite{Aartsen:2013jdh,Aartsen:2013bka,Kopper:2015vzf}. If at all these
three PeV events are from the GR, they
must be from the leptonic decay mode of the
GR\cite{Barger:2012mz}. However, there is no convincing reason why such a
sub-dominant decay mode is more favorable than the dominant hadronic
mode. Most probably, these three PeV events are not
from the GR, but are from the charged current or neutral current interactions
of the electron and/or tau neutrinos with the nuclei in ice.
An extremely high-energy muon neutrino with energy $\sim 8.7$ PeV was also reported
by the IceCube team\cite{Aartsen:2016xlq}. This track event deposited 2.6 PeV energy in the detectors.
The probabilities that the primary neutrino was a $\nu_\mu$, $\nu_\tau$, and ${\bar\nu}_{e}$
are 87.7\%, 10.9\%, and 1.4\%, respectively. Not only the probability of being $\bar\nu_e $ is low, 
but the inferred energy (8.7 PeV) is also different from the GR energy (6.3 PeV). So this event
is also likely not a GR event. 

\section{The potential ``missing Glashow resonance" problem}

If the three (1-2) PeV and one 8.6 PeV IceCube neutrinos are not from the GR, in a
model-independent way, one may argue that there could be a ``missing
GR'' problem from the IceCube data. This can be manifested as
follows. In astrophysical environments, neutrinos are produced via the
$p\gamma$ interactions
\be
p+\gamma \rightarrow \Delta^+\rightarrow  
 \left\{ 
\begin{array}{l l}
 p\,\pi^0 
\rightarrow p\gamma\gamma
, & \quad 
\\  n\,\pi^+  
\rightarrow n\mu^+\nu_{\mu} \rightarrow n e^+\,\nu_e \, {\bar\nu_{\mu}}\,\nu_{\mu},
& \quad  
\\
\end{array} \right. 
\label{pgamma}
\ee
\be
p+\gamma \rightarrow X\,\pi^{\pm},
\label{pgamma-2}
\ee
or $pp$ interactions
\begin{eqnarray}
& pp\rightarrow X\,\pi^{\pm}, \nonumber \\
& \pi^+\rightarrow \mu^+ \,\nu_{\mu}\rightarrow e^+\,\nu_e \, {\bar\nu_{\mu}}\,\nu_{\mu}, \nonumber \\
& \pi^-\rightarrow \mu^- \,\bar\nu_{\mu}\rightarrow e^-\,\bar\nu_e \, {\nu_{\mu}}\,\bar\nu_{\mu}.
\label{pp}
\end{eqnarray}
In all the channels, one would expect that anti-neutrinos contribute
to $\sim 1/2$ of the total neutrino and anti-neutrino flux. Consider
vacuum neutrino oscillations that evenly distribute anti-neutrinos in
three species, the $\bar\nu_e$ flux is about 1/6 of the total neutrino
and anti-neutrino flux \cite{note0}.

Suppose that the three (1-2) PeV neutrinos (with average energy $\sim
1.4$ PeV) are produced by cosmic protons through the standard
$p\gamma$, $pp$ processes, one may estimate the expected number of 6.3
PeV GR $\bar\nu_e$'s by IceCube as
\be
 N_{\bar\nu_e,6.3{\rm PeV}}  \simeq  N_{\rm (1-2)PeV} \cdot
 \left(\frac{6.3}{1.4}\right)^{-p} \cdot 240 \cdot 68\% \cdot \frac{1}{6}.
\ee
The power index $p$ in the range $2$ to $3$ is the typical
value for cosmic ray spectrum between knee and ankle where multi-PeV
neutrinos/anti-neutrinos can be generated \cite{Hummer:2010ai,Aartsen:2016oji,Laha:2013eev}. For the above range of $p$
we obtain $1 \le N_{\bar\nu_e,6.3{\rm PeV}}\le 4$.
If the 8.3 PeV neutrino is from GR, then it is consistent with the
above estimate. However, if it is not from the GR, as suggested by its
small probability of being a $\bar\nu_e$ and the different energy from
6.3 PeV, then the expected GR neutrinos would be
\bary
 N_{\bar\nu_e,6.3{\rm PeV}} & \simeq & 1 \cdot
 \left(\frac{6.3}{8.7}\right)^{-p} \cdot 240\cdot 68\% \cdot
 \frac{1}{6}.
\eary
For the above range of $p$, the expected GR events will be in the range
$50 \le  N_{\bar\nu_e,6.3{\rm PeV}}\le 70$.
The above estimate raises a possible
``missing GR $\bar\nu_e$ problem'', especially if the 8.7 PeV event is not from the GR.
Some other models also suggested that at the resonance peak, the event rate of $pp-$ and
$p\gamma-$ generated GR $\bar\nu_e$ would be $\sim 3.2-3.5$ per year
and $0.6-0.8$ per year, respectively\cite{Biehl:2016psj,Xing:2011zm,Bhattacharya:2011qu,Bustamante:2016ciw,Shoemaker:2015qul,Vincent:2016nut}, 
already exceeding the non-detection limit with a large margin. Also
lack of GR resonance event is pointed out in Ref.\cite{Kistler:2016ask}.

\section{Constraints on neutrino-emission mechanism and environments with the missing GR $\bar\nu_e$}
Due to the large error boxes, the origin of high-energy neutrinos
detected by IceCube remain unknown. The proposed neutrino sources
include
blazars\cite{Murase:2014foa,Dermer:2014vaa,Padovani:2014bha,Sahu:2014fua,Miranda:2015ema,Padovani:2015mba,Padovani:2016wwn},
gamma-ray bursts (GRBs) \cite{Waxman:1997ti,Rachen:1998fd,Meszaros:2001ms,Dermer:2003zv,Razzaque:2004yv,Kashti:2005qa,Murase:2006mm,Gupta:2006jm,Murase:2013ffa,Senno:2015tsn,Anchordoqui:2014yva}, 
hypernovae\cite{Senno:2015tra,Liu:2013wia,He:2013cqa}, intergalactic 
shocks\cite{Loeb:2006tw,Murase:2008yt,Berezinsky:1996wx}, and starburst galaxies\cite{Murase:2013rfa,Xiao:2016rvd}, etc. A stringent constraint on the
associations of IceCube neutrinos with GRBs has been placed\cite{Abbasi:2012zw,Aartsen:2014aqy}, 
which posed interesting constraints on 
GRB models \cite{Zhang:2012qy,He:2012tq}. Possible associations of high-energy
neutrinos with blazars have been 
suggested\cite{Murase:2014foa,Dermer:2014vaa,Padovani:2014bha,Sahu:2014fua,Miranda:2015ema,Padovani:2015mba,Padovani:2016wwn,Kadler:2016ygj}, but the case
is not conclusive. Furthermore, it is unknown whether the $p\gamma$ or the
$pp$ interactions are the dominant channel to produce these
high-energy neutrinos.

If the ``missing GR $\bar\nu_e$ problem" indeed exists, interesting
constraints can be placed on the neutrino-generation mechanisms. In
order to suppress the $\bar\nu_e$ flux, one may draw three conclusions:
(1) $p\gamma$ rather than $pp$ is the dominant channel to produce
high-energy neutrinos; (2) $p\gamma$ interactions proceed in the $\Delta^+$-resonance
channel, with the multi-pion channels suppressed; (3) the magnetic field and photon energy
densities in the $p\gamma$ emission region is such that significant
$\mu^+$ cooling occurs before decaying, yet $\pi^+$'s essentially do not cool 
before decaying. 

The required $p\gamma$ dominance can be readily seen from
Eqs.(\ref{pgamma} -- \ref{pp}). The main difference is that
$p\gamma$ produces $\pi^+$ only at $\Delta^+$-resonance, while both multi-pion $p\gamma$
and $pp$ processes produce both $\pi^+$ and
$\pi^-$ \cite{note,note2,note3}. When $\pi^+$ decays, it produces
$\mu^+$ and $\nu_\mu$. No significant anti-neutrinos can be produced
before $\mu^+$ decays. For $\pi^-$, $\mu^-$ and $\bar\nu_\mu$ are
produced immediately after $\pi^-$ decay, so that some $\bar\nu_e$'s
would reach IceCube due to vacuum oscillation as $\bar\nu_\mu$
propagates towards Earth. Of course, when muons decay, anti-neutrinos
would in any case be produced. In the rest frame, pions and muons have
decay time scales
\be
\tau_\pi^0 \simeq 2.6\times 10^{-8}~{\rm s}
\ee
and
\be
\tau_\mu^0 \simeq 2.2\times 10^{-6}~{\rm s},
\ee
respectively. Since the muon lifetime is much longer than the pion
lifetime, it is possible to suppress high-energy anti-neutrino flux
through muon cooling. This is relevant only for $p\gamma$
interactions at $\Delta^+$-resonance. As a result, one may draw the conclusion that 
$p\gamma$ is the dominant channel to produce the neutrinos detected by IceCube
if one indeed has the ``missing GR $\bar\nu_e$ problem". 

Next, for $p\gamma$ interactions, one needs the multi-pion channel (Eq.(\ref{pgamma-2})).
The $p\gamma$ interactions have a peak at the $\Delta^+$ resonance but a moderate
drop at higher photon energies, where multi-pion processes operate. To suppress 
multi-pion $p\gamma$ interactions, one would require a soft target photon spectrum
with rapid drop of photon flux at high energies. This is consistent with most models 
that invoke a synchrotron seed photon field as the targets, but disfavors the models
invoke a thermal photon seeds, such as the choked jet models for GRBs\cite{Waxman:1997ti,Rachen:1998fd,Meszaros:2001ms,Dermer:2003zv,Razzaque:2004yv,Kashti:2005qa,Murase:2006mm,Gupta:2006jm,Murase:2013ffa,Senno:2015tsn,Anchordoqui:2014yva}.
This is consistent with the non-detection of neutrinos associated with GRBs\cite{Abbasi:2012zw,Aartsen:2014aqy}.

Finally, for $p\gamma$ interactions, one needs to produce several (1-2) PeV neutrinos but
suppress $\sim$ 6.3 PeV anti-neutrinos. One may argue that there might be an intrinsic
cutoff in the injected cosmic-ray spectrum in this energy range, as is expected in
some models\cite{Waxman:1997ti,Rachen:1998fd,Meszaros:2001ms,Dermer:2003zv,Razzaque:2004yv,Kashti:2005qa,Murase:2006mm,Gupta:2006jm,Murase:2013ffa,Senno:2015tsn,Anchordoqui:2014yva}. However, the detection of the 8.7 PeV event (likely $\nu_\mu$,
which can be directly produced via $\pi^+$ decay)
disfavors such a possibility (again assuming that the event is not due GR). One is therefore
left with the following possibility, i.e.
\bary
 t_{\rm \pi,c} & > & \tau_\pi ~~~~~ {\rm for~2~PeV ~neutrinos},  \label{tpic} \\
 t_{\rm \mu,c} & < & \tau_\mu ~~~~~ {\rm for~6.3~PeV ~neutrinos} , \label{tmuc}
\eary
where $t_{i,c}$ is the cooling time scale and $\tau_i = \gamma_i
\tau_i^0$ is the lifetime of the species $i$ (for $\tau^\pm$ and
$\mu^\pm$, respectively).

In general, the energy loss rate of a high energy particle with
Lorentz factor $\gamma_i$ (and corresponding dimensionless speed
$\beta_i$) reads
\be
| {\dot E_i} |= \frac{4}{3} \sigma_{T,i} c\,\beta^2_{i}
  \gamma^2_{i} U_T,
\ee
where $i=\pi^{\pm}$, $\mu^{\pm}$ and 
$\sigma_{T,i}$ is the Thomson scattering cross section for particle
$i$, which can be calculated from $\sigma_{T,i}=({m_e}/{m_i})^2
\sigma_T$, with $\sigma_T\simeq 6.65\times 10^{-25}\, cm^2$ being the
Thomson cross section for electrons. Since the particles are
relativistic, one has $\beta_{i}\simeq 1 $
and $\gamma_{i}=E_{i}/m_{i}c^2$. Considering both synchrotron and
inverse Compton cooling, the total energy density in the emission region is defined by
\be
U_T=U_B+U_{ph} =\frac{B^2}{8\pi} (1+Y) + U_{E,ph},
\ee
where $B$ is the magnetic field strength in the emission region, $Y
\sim 1$ is the Compton parameter for synchrotron self-Compton (SSC)
process, and $U_{E,ph}$ is the external photon energy density in the
emission region. 
The radiative cooling time for the particle $i$ is given by
\be
t_{i,c}=\frac{E_i}{| {\dot E_i} |}=\frac{3}{4} 
 \frac{m_i c}{\gamma_i\sigma_{T,i} U_T}.
\label{tcooli}
\ee

In a photohadronic process, the pion carries
approximately 20\% of the UHE proton energy, whereas, in the pion
decay $\pi^{\pm}\rightarrow
\mu^{\pm}\,\nu_{\mu}(\bar\nu_{\mu})$ the muon carries about 75\% of the
pion energy. In the decay of the charged pion to leptons, 
each lepton carries about 25\% of the pion
energy. So in a photohadronic process about 5\% of the proton energy
is taken away by a single neutrino. If a GR event with 6.3 PeV energy $\bar\nu_e$ is observed in
IceCube from the photohadronic process, it corresponds to a parent UHE proton
energy $E_p\simeq 127$ PeV, pion energy  $E_{\pi}\simeq
25$ PeV and muon energy $E_{\mu}\simeq 19$ PeV respectively. Given $m_\pi \sim 139.57$ MeV,
and $m_\mu \sim 105.66$ MeV, the estimated Lorentz factors of 
pions and muons are $\gamma_{\pi} \simeq \gamma_{\mu}\simeq 1.8\times 10^8$.
To be able to produce a 2 PeV $\nu_\mu$ from $\pi^+$ decay, the required pion Lorentz factor is $\gamma_{\pi}
\simeq 5.7\times 10^7$.

The conditions (\ref{tpic}) and (\ref{tmuc}) then lead to a constraint on the total energy density
\be
\frac{3}{4} \frac{m_\mu c}{\gamma_\mu^2 \sigma_{T,\mu} \tau^0_\mu} < U_T < 
\frac{3}{4} \frac{m_\pi c}{\gamma_\pi^2 \sigma_{T,\pi} \tau^0_\pi},
\ee
where $\gamma_\mu \simeq 1.8\times 10^8$, $\gamma_\pi \simeq 5.7\times 10^7$. This gives
\be
 3.8\times 10^3~{\rm erg/cm^3} < U_T < 7.3 \times 10^6 ~{\rm erg/cm^3}.
 \label{constraint}
\ee
In the case of synchrotron cooling dominated sources so that $U_T
\simeq U_B = B^2/8\pi$, this condition 
can be expressed as a constraint on the magnetic field strength in the
source
\be
 310~{\rm G} < B < 1.4 \times 10^4 ~{\rm G}.
 \label{B-constraint}
\ee

These constraints have important implications on the astrophysical
sources of high-energy neutrinos. The co-moving frame magnetic field
strength of a relativistic jet with wind luminosity $L$, bulk Lorentz
factor $\Gamma$, and magnetic fraction parameter $\epsilon_B = L_B / L
< 1$ can be estimated as
\begin{eqnarray}
B' & = & \left(\frac{2 \epsilon_B L}{\Gamma^2 r^2 c}\right)^{1/2} \nonumber \\
 & \simeq & (2.6 \times 10^4~{\rm G}) L_{52}^{1/2} \epsilon_{B}^{1/2} \Gamma_{2.5}^{-1} r_{14}^{-1}  \\
 & \simeq & (8.2\times 10^2~{\rm G}) L_{48}^{1/2} \epsilon_{B}^{1/2} \Gamma_{1}^{-1} r_{15}^{-1},
\end{eqnarray}
where the convention $Q=10^n Q_n$ in cgs units and the characteristic
parameters for GRBs and blazars have been adopted. One can immediately
see that a GRB internal shock environment roughly satisfies this
constraint. However, since so far no IceCube neutrinos have been
detected to be associated with GRBs\cite{Abbasi:2012zw,Aartsen:2014aqy}, it suggests that
successful GRBs are not likely the dominant sources for the IceCube
neutrinos. For blazars, in order to satisfy the constraint, one needs
to reduce the emission radius to $r \sim 10^{15}$ cm, which is about
100 times of the BH Schwarzschild radius, suggesting a core origin of
neutrinos. Alternatively, one may have a large emission radius, but the
neutrino emission region should be permeated with external photons with energy
density satisfying the constraint (\ref{constraint}). Most other
high-energy neutrino models (hypernova or intergalactic shocks)\cite{Senno:2015tra,Liu:2013wia,He:2013cqa,Loeb:2006tw,Murase:2008yt,Berezinsky:1996wx}
typically have a much weaker
magnetic field strength in the emission region. These models would
work only when the constraint (\ref{constraint}) is satisfied via an
in-situ photon background.

\section{Conclusions} 
We argue that with the current data, there
might be a ``missing GR problem". The case is only
marginal if the 8.7 PeV neutrino detected by IceCube is a GR event, the probability
of which is low. If this event is not a GR, then the missing GR problem is very severe,
and some interesting constraints on the origin of IceCube
high-energy neutrinos can be placed. The neutrino production mechanism
is likely $p\gamma$ rather than $pp$. For $p\gamma$ processes, the interactions
should mostly proceed at the $\Delta^+$-resonance, with the multi-pion interactions
suppressed. Also, the neutrino emission site should have a significant amount of magnetic
field or photon energy density so that $\mu^+$ can significantly cool
before decaying and producing anti-neutrinos. In the meantime, the
energy density should not be too high to cool $\pi^+$, so that 2 PeV
neutrinos can be generated. This condition places an interesting
constraint (Eq.(\ref{constraint})) that any high-energy neutrino model
has to satisfy. \\

This work is partially supported by DGAPA-UNAM (Mexico) 
Project No. IN110815 (S.S.) and by
NASA through grants NNX15AK85G and NNX14AF85G (B.Z.). S.S. is a Japan
Society for the Promotion of Science (JSPS) invitational fellow. We thank Kohta Murase, Shigehiro Nagataki, Ranjan Laha and Matt Kistler for important comments on the paper.




\begin{thebibliography}{55}
\bibitem{note0}
Neutrino oscillations in matter do not play a significant role in most
high-energy neutrino emission models, so that their effect is
neglected in our analysis.
\bibitem{note}
A small amount of  $\pi^-$ can be produced from the multi pion production through $p\gamma\rightarrow n\pi^+ + n_{\pi}(\pi^+\pi^-)$, where $n_{\pi}$ is the pion multiplicity factor. In the source, the back reaction $n\gamma\rightarrow p\pi^-$ could also take place depending on the optical depth of the source. 
In any case, the amount of $\bar\nu_e$ is negligibly small, so that $N_{\nu_{\mu}}=N_{{\bar\nu}_{\mu}} =N_{\nu_e}  \gg N_{{\bar\nu}_e}$ for $p\gamma$ interactions.
\end{thebibliography}



\end{document}